\documentclass[namedreferences,hyperref,optionalrh]{spr-sola}
\usepackage{fix-cm}
\usepackage{graphicx}        
\usepackage{color}           
\usepackage[dvipsnames]{xcolor}
\usepackage[normalem]{ulem}
\usepackage{xspace}
\usepackage{epstopdf}	

\def \cesamxx{Cesam2k20\xspace}

\newcommand{\deriv}[2]{\frac{{\mathrm d} #1}{{\mathrm d} #2}}

\chardef\us=`\_

\begin{document}

\begin{frontmatter}
\title{The impact of the transport of chemicals and electronic screening on helioseismic and neutrino observations in solar models}

\author[addressref={aff1},corref,email={morgan.deal@umontpellier.fr}]{\inits{M.}\fnm{Morgan}~\snm{Deal}\orcid{0000-0001-6774-3587}}
\author[addressref=aff2,email={gbuldgen@uliege.be}]{\inits{G.}\fnm{Gaël}~\snm{Buldgen}\orcid{0000-0001-6357-1992}}
\author[addressref=aff3]{\inits{L.}\fnm{Louis}~\snm{Manchon}\orcid{0000-0002-0104-6444}}
\author[addressref={aff3,aff4}]{\inits{Y.}\fnm{Yveline}~\snm{Lebreton}}
\author[addressref=aff2]{\inits{A.}\fnm{Arlette}~\snm{Noels}}
\author[addressref=aff2]{\inits{R.}\fnm{Richard}~\snm{Scuflaire}\orcid{0009-0000-8017-7962}}
\address[id=aff1]{LUPM, Universit\'e de Montpellier, CNRS, Place Eug\`ene Bataillon, 34095 Montpellier, France}
\address[id=aff2]{STAR Institute, Université de Liège, Liège, Belgium}
\address[id=aff3]{LIRA, Observatoire de Paris, Université PSL, Sorbonne Université, Université Paris Cité, CY Cergy Paris Université, CNRS, 92190 Meudon, France}
\address[id=aff4]{Université de Rennes, CNRS, IPR (Institut de Physique de Rennes) -- UMR 6251, 35000 Rennes, France}

\runningauthor{Deal et al.}
\runningtitle{\textit{Solar Physics} Example Article}

\begin{abstract}
The transport of chemical elements in stellar interiors is one of the greatest sources of uncertainties of solar and stellar modelling. The Sun, with its exquisite spectroscopic, helioseismic and neutrino observations, offers a prime environment to test the prescriptions used for both microscopic and macroscopic transport processes. We study in detail the impact of various formalisms for atomic diffusion on helioseismic constraints in both CLES \citep{scuflaire08} and \cesamxx (\citealt{morel08,marques13,deal18}) models and compare both codes in detail. Moreover, due to the inability of standard models using microscopic diffusion to reproduce light element depletion in the Sun (Li, Be), another efficient process must be included to reproduce these constraints (rotation-induced: \citealt{eggenberger22}, overshooting -or penetrative convection- below the convective envelope: \citealt{thevenin17}, or ad hoc turbulence: \citealt{lebreton87,richer00}). However, introducing such an extra mixing leads to issues with the CNO neutrino fluxes (see \citealt{buldgen23}), which seem to be systematically lower than the Borexino observations \citep{borexino22}. Another key aspect to consider when reconciling models with neutrino fluxes is the impact of electronic screening \citep{mussack11}.
\end{abstract}
\end{frontmatter}

\section{Introduction}
     \label{S-Introduction} 

The Sun is a key calibrator in stellar physics. As such, it also serves as a test case for the robustness of stellar evolution code and any new revision of physical processes to be implemented on a wider scale. In the era of space-based asteroseismology (CoRoT: \citealt{CoRoT}, \textit{Kepler:} \citealt{kepler}, TESS: \citealt{TESS}) and looking forward to the launch of the PLATO mission by the end of 2026 \citep{rauer25}, it is worthwhile to analyze in detail the existing differences between solar-calibrated models computed using various stellar evolution codes. Our main motivation is to follow the seminal work of \citet{Boothroyd2003} that looked into the details of potential uncertainties that might affect the conclusions of comparisons of solar models with the exquisite quality of spectroscopic, helioseismic and neutrino flux constraints. Twenty-two years after their work, the precision of the available constraints has improved further, while the uncertainties in stellar models have remained, ultimately, quite similar, with asteroseismic observations further showcasing the limitations of stellar evolutionary models \citep[see e.g.][]{LGM14a,LGM14b,farnir20}. The Sun is no stranger to such limitations, as standard solar models still lack a full depiction of dynamical processes that would reproduce light element depletion, and there is still a debate about the origin of the observed differences in sound speed, with recent work questioning the abundances from \citet{asplund21} while mounting evidence seems to point to shortcomings in opacity computations \citep[see][for an extensive review on solar modelling]{JCD2021}.

In this paper, we follow the work of \citet{Boothroyd2003} and the CoRoT ESTA exercises \citep{lebreton07} that provided detailed comparisons of stellar models, focusing here on the Sun. We compare results of the \cesamxx and the CLES stellar evolution codes, which have both been used in detailed helioseismic comparisons and differ in their implementation of key physical phenomena: transport and energy production. The models, calibration and inversion methods, and a reference comparison model (without atomic diffusion) are presented in Section 1. The comparisons of both CLES and \cesamxx for Standard Solar Models are presented in Section 2, together with additional tests on the atomic diffusion formalism. The adjustment of the lithium surface abundance with an additional turbulent mixing is discussed in Section 4. The contribution of electronic screening from nuclear reactions and the neutrino fluxes are presented in Section 5. The inversion results for all models are discussed in Section 6, including a specific discussion on the convection formalism, and we conclude on all the comparisons in Section 7. 

\section{Solar models}

The impact of different processes on solar modelling is tested with two stellar evolution codes, the Code Liègeois d'Evolution Stellaire (CLES, \citealt{scuflaire08}) and the public French Code d'Evolution Stellaire Adaptatif et Modulaire (\cesamxx\footnote{\url{https://www.ias.u-psud.fr/cesam2k20/}}, \citealt{morel08,marques13,deal18}). Both codes include similar input physics, which allows for useful comparisons, and additional processes are also tested with \cesamxx. Both codes have already shown that they provide very similar results within the ESTA / CoRoT project \citep{monteiro06,Montalban2007,lebreton07,lebreton08,montalban08} at the level of precision required to exploit asteroseismic data. However, when working with precision and accuracy requirements of helioseismic applications, more in-depth comparisons of both physical ingredients and numerical procedures are required, as variations of $1\%$ in local thermodynamic quantities, which would be overlooked in asteroseismology, must now be considered highly significant.

\subsection{Input physics}

All models are computed with the version 7 of the SAHA-S equation of state \citep{gryaznov04,baturin13} (except for two models, see Sec.~\ref{sec:conv}) and OPAL opacity tables for high temperature \citep{iglesias96} complemented at low temperature by the Wichita tables from \cite{ferguson05}. The choice of SAHA-S is driven by the fact the equation of state has been extensively tested for its resolution and its interpolation routines \citep{baturin17} and it has been shown to be better suited for Standard Solar Models than OPAL \citep{Vorontsov2013}. Moreover, it is one of the only equations of state in common between both stellar evolution codes. The mixture of heavy elements follows photospheric \cite{asplund09} (hereafter AGSS09), except for lithium and beryllium for which the meteoritic abundances are used. Nuclear reaction rates are taken from the NACRE compilation \citep{angulo99}, except for the $^{14}\mathrm{N}(p,\gamma)^{15}\mathrm{O}$ reaction, for which we adopted the LUNA rate \citep{imbriani04}. The atmosphere boundary condition follows the $T(\tau)$ relation of \citet[Model C]{vernazza81}. All models are computed from the pre-main sequence (PMS). The other (or changed) input physics are listed in Table~\ref{tab:mod} for each model and are detailed in the following subsections. We do not consider the effects of the solar wind, as there are intrinsic differences in the implementation of mass loss between the two codes and the impact of the solar wind depends largely on its composition \citep{Boothroyd+Sackmann03,Wood+18,Zhang+19}, which cannot be easily adjusted. We note that such comparisons would be relevant, particularly in the context of solar models taking into account proto-solar accretion based on planetary formation \citep{Kunitomo+Guillot21,Kunitomo+22}.

\subsection{Convection}\label{sec:conv}

The mixing length theory (MLT) is the best known \textit{ad hoc} approach to modelling convection in stellar evolution codes. It assumes that the whole convective flux is carried by the largest mode of turbulence (hence the turbulence spectrum is reduced to a Dirac distribution). This mode has a length $\ell_{\rm MLT}$ which is related to the local pressure scale height $H_{\rm p}$ by the proportionality constant $\alpha_{\rm MLT}$. This is the main free parameter of the MLT, but not the only one: others are used, e.g. to control the shape of the convective bubbles, but they have less impact on the stellar structure and evolution. The \citet[][hereafter CGM]{canuto96} formalism belongs to the category of Full Spectrum of Turbulence (FST) models and assumes a Kolmogorov spectrum for the turbulence in the convective zones \citep{canuto91}. Similarly to MLT, CGM is based on a free parameter $\alpha_{\rm CGM}$ that relates a characteristic length of the convection to local properties of the medium. CGM proposed to define this length as $\ell_{\rm CGM} = z + \alpha_{\rm CGM} H_{\rm p, top/bot}$, where $z$ is the distance to the nearest boundary and $H_{\rm p, top/bot}$ is the height of the pressure scale at the nearest boundary. The term $\alpha_{\rm CGM} H_{\rm p, top/bot}$ should be understood as the penetration length in the adjacent zone. Following \citet{heiter02}, the implementation in \cesamxx \citep{samadi06} uses the same definition for $\ell_{\rm CGM}$ as in the MLT: $\ell_{\rm CGM} = \alpha_{\rm CGM} H_{\rm p}$. The convective penetration (not considered here) can then be controlled by another free parameter.

The entropy-calibration method \citep[][hereafter ECM]{spada18,spada19,spada21,manchon24} was designed to put a tight constraint on the free parameter $\alpha$ of MLT or CGM. In the adiabatic region, the behaviour of the convection does not depend on the chosen convection paradigm, but the value of $\alpha$ determines the value of the adiabat \citep{gough76}. What value this adiabat should have cannot be determined from physical principles. However, it can be calculated with 3D simulations of surface convection \citep[see, e.g.,][]{ludwig99,magic15,tanner16}. From large sets of such 3D simulations one can derive prescriptions for the entropy of the adiabat as a function of the global parameters $T_{\rm eff}$, $\log g$ and $[\rm Fe/H]$ of a star (for FGK stars). Using these in \cesamxx, one can adjust the $\alpha$ parameter along the evolution so that the adiabat of the model corresponds to the prescribed value for the same global parameters. 
It should be stress that, except for the varying $\alpha$, EC formalism is identical to MLT formalism. Therefore, no improvement of the modelling of the surface layers should be expected when using ECM in place of MLT. However, it allows one to have an $\alpha$ parameter that varies with time and does not need to be prescribed in any other way.

Most models of this study are computed following the MLT \citep[][]{bohm58,Cox1968} with a constant solar calibrated $\alpha_\mathrm{MLT}$ parameter. No convective overshoot is included for these models. We also computed a \cesamxx model with the formalism of CGM (CESAM CGM) and another one with the entropy calibration (CESAM ECM). The ECM model is computed with the OPAL equation of state \citep{rogers02} because the entropy is not an output of the current implementation of the SAHA-S equation of state in \cesamxx.

\subsection{Transport of chemical elements}

In both codes, the transport of chemical elements is taken into account by solving the diffusion equation
\begin{equation}
\rho\frac{\partial X_i}{\partial t}=\frac{1}{r^2}\frac{\partial}{\partial r}\left[r^2\rho {D_\mathrm{turb}}\frac{\partial X_i}{\partial r}\right]-\frac{1}{r^2}\frac{\partial}{\partial r}[r^2\rho v_{i}] + A_i m_p \left[\sum_j (r_{ji} -  r_{ij}) \right],
\end{equation}
\noindent where $X_i$ is the mass fraction of element $i$, $A_i$ its atomic mass, $v_i$ its atomic diffusion velocity, $\rho$ the density in the considered layer, $D_\mathrm{turb}$ the turbulent diffusion coefficient, $m_p$ the mass of
a proton, and $r_{ij}$ the reaction rate (cm$^{-3}$.s$^{-1}$) of the reaction that transforms element $i$ into $j$.

\subsubsection{Atomic diffusion}\label{sec:diff}

Atomic diffusion is taken into account in both codes with different formalisms.

\textit{CLES:} While nuclear reactions are accounted for individual elements, the atomic diffusion velocities $v_i$ are computed for hydrogen (X), helium (Y) and an average metal (Z) using the \cite{thoul94} formalism. By default, the screening potentials of \citet{Paquette1986} are included in the formalism instead of the Debye sphere cutoff applied in the classical \citet{thoul94} method. The use of the \citet{Paquette1986} coefficients became the default only in 2017 in CLES, with a first publication comparing the prescriptions in the code in 2019 \citep{Buldgen2019Frontiers}.

\textit{\cesamxx:} Atomic diffusion velocities $v_i$ are individually computed for H, He, Li, Be, B, C, N, O, Ne, Na, Mg, Al, Si, S, Ca and Fe (and some isotopes) using either \cite{michaud93} or \cite{burgers69} formalism. The latter is computed with the screening potentials of \citet{Paquette1986}. Moreover, radiative accelerations are taken into account for one model following the Single Valued Parameters (SVP) approximation \citep{alecian20}.

\subsubsection{Turbulent mixing}\label{sec:turb}

Ad hoc turbulent mixing is included in some models to account for the transport induced by macroscopic processes such as rotation-induced mixing. One way to model it is to use the expression proposed by \cite{richer00} where the turbulent diffusion coefficient is defined as
\begin{equation}\label{eq:dturb1}
    D_\mathrm{turb,add}=\omega D_{\mathrm{He},0} \left(\frac{\rho_0}{\rho}\right)^n,
\end{equation}
\noindent where $\omega$ and $n$ are constant. The index "$0$" corresponds to a reference depth in the star, $D_\mathrm{He,0}$ and $\rho_0$ are the diffusion coefficient of helium and the density at this reference temperature, respectively. The reference depth can be defined either in temperature (DT0 models) with $\omega=400$ and $n=3$ \citep[see e.g.][]{richard02,semenova20,dumont21} or in mass (DM0 models) with $\omega=10000$ and $n=4$ \citep[see e.g.][]{richard01,michaud11}.

Another way is to use the expression of \cite{proffitt91} defined as 
\begin{equation}\label{eq:dturb2}
    D_\mathrm{turb,add}=\omega \left(\frac{\rho_\mathrm{cz}}{\rho}\right)^n,
\end{equation}
\noindent where $\rho_\mathrm{cz}$ is the density at the bottom of the convective envelope. The exponent $n=3$ is adopted for such DCZ models (similarly to \citealt{proffitt91}).

\subsection{Electronic screening}\label{Sec:Screening}

The treatment of the screening of nuclear reactions has been questioned by various authors over the years \citep[see][for a recent discussion]{Dappen2024}, with various formalisms being available in stellar evolution codes. So far, all of the implemented formalisms are static in nature, meaning that they do not consider the impact of velocity effects on the potential barrier that two nuclei have to overcome. One may consider so-called ``weak'' \citep{Salpeter1954}, ``intermediate'' \citep{Graboske1973} or ``strong'' \citep{Mitler1977} screening, but none of these formalisms take dynamical effects into account. Early works by \citet{Shaviv1996} questioned the validity of this static approach and showed with molecular dynamics simulations that dynamical effects should be considered. \citet{Brown1997} carried out analytical plasma effects calculations that essentially reduced to a validation of the \citet{Salpeter1954} formula while \citet{Mao2009} and \citet{mussack11} confirmed the results of the molecular dynamics simulations of \citet{Shaviv1996}, \citet{Shaviv2004}. So far, the controversy remains open \citep{Dappen2024} and it is therefore worth testing the impact of electronic screening on the predictions of solar models, especially given that it was historically studied to investigate the solar neutrino problem \citep{Dzitko1995}. Works by \citet{Fiorentini2001} and \citet{Weiss2001} concluded that screening was required to reproduce the solar sound speed profile as determined from helioseismology, while \citet{mussack11} show a slight improvement of the sound speed profile of their model including dynamical screening effects.

We follow the recommendations of \citet{mussack11} regarding the impact of screening on the nuclear reaction rates, namely that using dynamical screening corrections for the pp chain is the same as reducing the efficiency of the nuclear reaction rates by $\approx 5\%$ compared to a case with a weak screening. We apply this correction to all nuclear reactions while keeping the screening active. This approach is approximative, as screening effects are thought to be more important for the CNO cycle reaction than for the pp chain. We also carry out a test where the screening expression is simply entirely removed from all reactions in CLES.

\begin{table}
\caption{List of models computed with the AGSS09 mixture. \textit{no diff} stands for no atomic diffusion, \textit{SSM} stands for Standard Solar Model, \textit{Burgers} stands for models including the \cite{burgers69} formalism for atomic diffusion (B69), \textit{SVP} stands for the Single Valued Parameters for radiative accelerations computations \citep{alecian20}. \textit{DT0}, \textit{DM0}, \textit{DCZ} stand for the different additional mixing prescriptions with a reference depth of the mixing in temperature and mass \citep[R00][]{richer00}, and base of the convective envelope \citep[PM91][]{proffitt91}, respectively. \textit{nuc} stands for models with a reduced efficiency of $5\%$ of the nuclear rates and \textit{noscreen} stand for a model without nuclear screening. \textit{CGM} stands for a model including the \cite{canuto96} formalism for convection, and \textit{ECM} for Entropy Calibrated Model. \textit{OPAL} indicates that the model is computed with the OPAL equation of state instead of the SAHA-S one. MP93 and T94 stand for \cite{michaud93} and \cite{thoul94}, respectively.
}
\label{tab:mod}
\begin{tabular}{lccccc}     
\hline                     
Model & Atomic diff. & Rad. acc. & $D_\mathrm{turb}$ & Conv. & Screening \\
\hline
CESAM nodiff & - & - & - & MLT & Y  \\
CLES nodiff & - & - & - & MLT & Y \\
CESAM SSM &  MP93 & - & - & MLT & Y  \\
CLES SSM & T94 & - & - & MLT & Y  \\
CESAM Burgers & B69 & - & - & MLT & Y   \\
CESAM SVP & MP93 & SVP & - & MLT & Y \\
CESAM DT0 & MP93 & - & R00 in T & MLT & Y \\
CLES DT0 & T94 & - & R00 in T & MLT & Y \\
CESAM DM0 & MP93 & - & R00 in M & MLT & Y \\
CLES DM0 & T94 & - & R00 in M & MLT & Y \\
CESAM DCZ & MP93 & - & PM91 & MLT & Y \\
CLES DCZ & T94 & - & PM91 & MLT & Y \\
CESAM nuc & MP93 & - & - & MLT & Y\\
CLES nuc & T94 & - & - & MLT & Y \\
CLES noscreen & T94 & - & - & MLT & N \\
CESAM CGM & MP93 & - & - & CGM & Y\\
CESAM SSM OPAL & MP93 & - & - & MLT & Y\\
CESAM ECM OPAL & MP93 & - & - & ECM & Y\\
\hline
\end{tabular}
\end{table}

\subsection{Optimisation procedure}

Solar models are calibrated using a Levenberg-Marquardt algorithm. For \cesamxx, the Optimal Stellar Model (OSM) code is used (see \citealt[][and references therein]{manchon24} for more details). A similar minimization procedure (called ``min-cles'') is used for CLES \citep[See][for a description and examples]{Buldgen2019}.

Solar calibrations are performed on the initial chemical composition ($Y_0$ and $(Z/X)_0$) and the mixing length parameter $\alpha_\mathrm{MLT}$. For the ECM, $\alpha_\mathrm{MLT}$ is replaced by an envelope overshoot parameter in order to achieve the require precision on the calibration. This compensates for the fact that the mixing length parameter is not a free parameter for this model and keeps the number of free parameters equal to the number of constraints. However, without this extra constraint, the solar calibrated ECM already has a precision of $10^{-3}$ for $R$ and $L$. The mass is fixed at $1$~M$_\odot$ and the age at $4.570$~Gyr. The constraints are the radius, the luminosity and $Z/X=0.0181$ (according to \citealt{asplund09} solar mixture), with a precision of $10^{-5}$. The observed solar values are taken from the IAU 2015 resolution B3 \citep{Mamajek2015} for both calibration procedures. When turbulent mixing is included in the models, the reference depth is calibrated (in temperature or mass) or the value of $\omega$ when the \cite{proffitt91} formalism is used. In these cases, we consider the additional constraint on the lithium surface abundance $A(\rm{Li})=0.96\pm0.05$~dex \citep{wang21}. The results of the calibrations are given in Table~\ref{tab:inputs}.

\begin{table}
\caption{Inferred input parameters and properties of the calibrated solar models. The lithium and beryllium abundances are provided in the scale $\log(N_\mathrm{Li/Be}/N_\mathrm{H}) + 12 $ with $N$ the number density of the element.} The lithium and beryllium abundances are not available for the CESAM Burgers models.
\label{tab:inputs}
\begin{tabular}{lccccccc}     
\hline                     
Model & $Y_0$ & $Z_0$ & $(Z/X)_0$  & $R_\mathrm{cz}$ [$R_\odot$] & $Y$ & $^7$Li & $^9$Be\\
\hline
CESAM nodiff & $0.2569$ & $0.01321$ & $0.0181$ & $0.7413$ & $0.2569$ & $3.11$ & $1.30$ \\
CLES nodiff & $0.2581$ & $0.01319$ & $0.0181$ & $0.7387$ & $0.2581$ & $2.97$ & $1.30$ \\
CESAM SSM & $0.2658$ & $0.01503$ & $0.0209$ & $0.7264$ & $0.2364$ & $2.88$ & $1.29$ \\
CLES SSM & $0.2670$ & $0.01493$ & $0.0208$ & $0.7249$ & $0.2383$ & $2.80$ & $1.34$  \\
CESAM Burgers & $0.2657$ & $0.01503$ & $0.0209$ & $0.7266$ & $0.2364$ & $-$ & $-$ \\
CESAM SVP & $0.2654$ & $0.01496$ & $0.0208$ & $0.7265$ & $0.2360$ & $2.88$ & $1.29$ \\
CESAM DT0 & $0.2601$ & $0.01410$ & $0.0194$ & $0.7304$ & $0.2435$ & $0.96$ & $1.11$ \\
CLES DT0 & $0.2625$ & $0.01420$ & $0.0196$ & $0.7287$ & $0.2469$ & $0.97$ & $1.12$ \\
CESAM DM0 & $0.2603$ & $0.01413$ & $0.0195$ & $0.7302$ & $0.2432$ & $0.96$ & $1.18$ \\
CLES DM0 & $0.2627$ & $0.01423$ & $0.0197$ & $0.7286$ & $0.2465$ & $0.98$ & $1.20$ \\
CESAM DCZ & $0.2604$ & $0.01414$ & $0.0195$ & $0.7302$ & $0.2431$ & $0.96$ & $1.13$ \\
CLES DCZ & $0.2626$ & $0.01424$ & $0.0197$ & $0.7285$ & $0.2464$ & $0.97$ & $1.14$ \\
CESAM nuc & $0.2651$ & $0.01510$ & $0.0210$ & $0.7286$ & $0.2349$ & $2.89$ & $1.29$ \\
CLES nuc & $0.2637$ & $0.01501$ & $0.0208$ & $0.7250$ & $0.2351$ & $2.79$ & $1.34$ \\
CLES noscreen & $0.2659$ & $0.01502$ & $0.0209$ & $0.7275$ & $0.2364$ & $2.86$ & $1.34$ \\
CESAM CGM & $0.2659$ & $0.01503$ & $0.0209$ & $0.7265$ & $0.2364$ & $2.94$ & $1.29$ \\
CESAM SSM OPAL & $0.2652$ & $0.01503$ & $0.0209$ & $0.7257$ & $0.2359$ & $2.88$ & $1.29$ \\
CESAM ECM OPAL & $0.2641$ & $0.01484$ & $0.0206$ & $0.7263$ & $0.2364$ & $2.99$ & $1.29$ \\
\hline
\end{tabular}
\end{table}

\subsection{Inversions of solar structure}

One of the most common quality tests for solar models are comparisons with the internal solar structure as measured from helioseismic data using inversion techniques. To do this, we computed the adiabatic oscillation frequencies and eigenfunctions for the models in Table \ref{tab:inputs} using the Liège adiabatic OScillation Code \citep[LOSC,][]{Scuflaire2008LOSC}.

Usually, models are compared using their adiabatic sound speed profiles, expressed as $c^{2}=\frac{\Gamma_{1}P}{\rho}$, with $P$ the pressure, $\rho$ the density and $\Gamma_{1}$ the first adiabatic exponent. This inversion usually serves as a basic quality check of the agreement with helioseismic data. Moreover, this test may also serve in this study as a form of resolution limit for inversion techniques regarding how far we can push their diagnostic potential, as well as an efficient tracking of potential disagreements between the models. In this work, we carry out both inversion of the squared adiabatic sound speed and the Ledoux discriminant. The latter is defined as
\begin{equation}
A=\frac{1}{\Gamma_{1}}\deriv{\ln P}{\ln r}-\deriv{\ln \rho}{\ln r},
\end{equation}
which is a proxy of the Brunt-Väisälä frequency and a direct measure of the stiffness of the stratification in the solar radiative zone. For both inversions, we use the classical Substractive Optimally Localized Averages (SOLA) technique \citep{Pijpers1994} that we applied extensively in previous works \citep[e.g.][]{Buldgen2019,Buldgen2024a,Buldgen2024b} and that has been widely used in helioseismology \citep[see e.g.][and references therein]{JCD2002,Thompson2003,Basu2008,JCD2021}. We use the solar frequency dataset of \citet{Basu2009} coupled with that of \citet{Davies2014}. We use to this end the usual linear variational formalism \citep{Dziembowski1990} that allows to write
\begin{equation}
\frac{\delta \nu^{n,l}}{\nu^{n,l}}=\int_{0}^{R}K^{n,l}_{s_{1},s_{2}}\frac{\delta s_{1}}{s_{1}}dr + \int_{0}^{R}K^{n,l}_{s_{2},s_{1}}\frac{\delta s_{2}}{s_{2}}dr + \frac{\mathcal{F}(\nu)}{Q^{n,l}} \label{eq:variational},
\end{equation}
with $K^{n,l}_{s_{i},s_{j}}$ the kernel function associated with the oscillation mode of frequency $\nu^{n,l}$ and $\mathcal{F}(\nu)$ a function depending only on frequency that is associated with the so-called surface effect, with $Q^{n,l}$ being the ratio of the mode energy to the energy of the radial mode of same frequency. The notation $\delta$ denotes a difference between an observed and a model quantity, defined as
\begin{equation}
\frac{x^{2}_\mathrm{Sun}-x^{2}_\mathrm{Model}}{x^{2}_\mathrm{Model}},
\end{equation}
where $x$ can be an oscillation frequency or the sound speed, density or Ledoux discriminant. The equation in the latter case will be slightly different as they will be written for $\delta A=A_\mathrm{Sun}-A_\mathrm{Model}$ and not for relative differences as is more common, see e.g. for sound speed, for which the structural perturbations are written as $\frac{\delta c^{2}}{c^{2}}$. The surface correction of the inversion is taken as the classical $6^\mathrm{th}$ order polynomial, and the parameters of the inversion are calibrated following \cite{Rabello1999}. The inversion results for the models studied here will be presented in Sect. \ref{Sec:InvRes}.

\subsection{Comparison of a solar model without atomic diffusion}\label{sec:nodiff}

\begin{figure}
\centering{\includegraphics[width=0.6\textwidth,clip=]{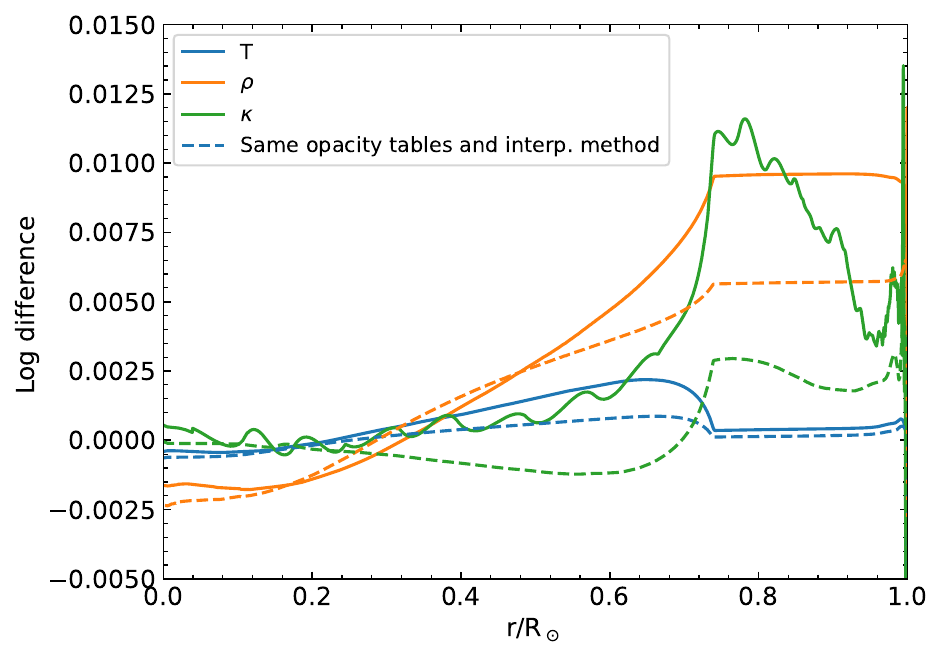}}
\small
        \caption{Logarithmic differences between models \cesamxx and CLES \textit{nodiff} models for different quantities as a function of the radius. The dashed lines represent the same quantities but the \cesamxx model is computed using the exact same opacity table and interpolation method as the CLES model.}
\label{fig:nodiff}
\end{figure}

The similarity between the solar models obtained with both codes is first tested on models without atomic diffusion in order to avoid any influence of the different treatment of this process. Logarithmic differences in the temperature, density and opacity profiles are shown in Fig.~\ref{fig:nodiff}. The differences are lower than $0.003$ for all profiles below $0.6$~R$_\odot$ and increase to $0.01$ above. We notice slight differences in the position of the base of the convective envelope ($R_\mathrm{cz}/R_\odot=0.7413$ for \cesamxx versus $0.7387$ for CLES) and in the initial helium abundance ($Y_0=0.2569$ against $0.2581$). We tested that the difference in helium comes from the treatment of $Z$ (mean metal in CLES for elements heavier than $^{17}$O and individual elements in \cesamxx up to iron). It induces a change in density that is compensated for by a change in helium initial abundance. To understand the origin of the observed differences in the position of the convective envelope between the two codes, we carried out the following tests. We selected a fixed radius in a \cesamxx model, chose the associated density, temperature, and chemical composition, and called the equation of state and opacity tables routines of CLES for these given thermodynamic coordinates. For the equation of state, no significant differences were observed for fixed coordinates. However, for opacities, CLES returned a a value $1.5\%$ larger than \cesamxx for a given $\rho$, $T$, $X$ and $Z$. As a test, we implemented in \cesamxx the exact same opacity table and the interpolation method implemented in CLES. The differences in the models are presented in dashed in Fig.~\ref{fig:nodiff}. We clearly see that a large part of the difference comes from the slightly different treatment of opacities in both codes. This may explain the differences in BCZ position and thus in density between the two codes, even without any transport processes included. Further detailed tests of the interpolation routines of both codes should be carried out to determine the exact origin of the observed differences. Overall, the comparison between the models obtained with both codes is satisfactory, and any major differences seen in the following comparison would come from the treatment of the processes specifically analyzed.

\section{Atomic diffusion}

Atomic diffusion refers to the microscopic transport of chemical elements induced by pressure, temperature, and composition gradients. It is mainly the competition between gravitational settling, that leads to a depletion of elements, and radiative accelerations that are induced by a transfer of momentum between photons and ions and push some elements toward the surface. Atomic diffusion has been shown to be a major ingredient of solar models, strongly improving, for example, the comparison with the sound speed profile inversions \citep{JCD1993}.

\begin{figure}
{\includegraphics[width=0.5\textwidth,clip=]{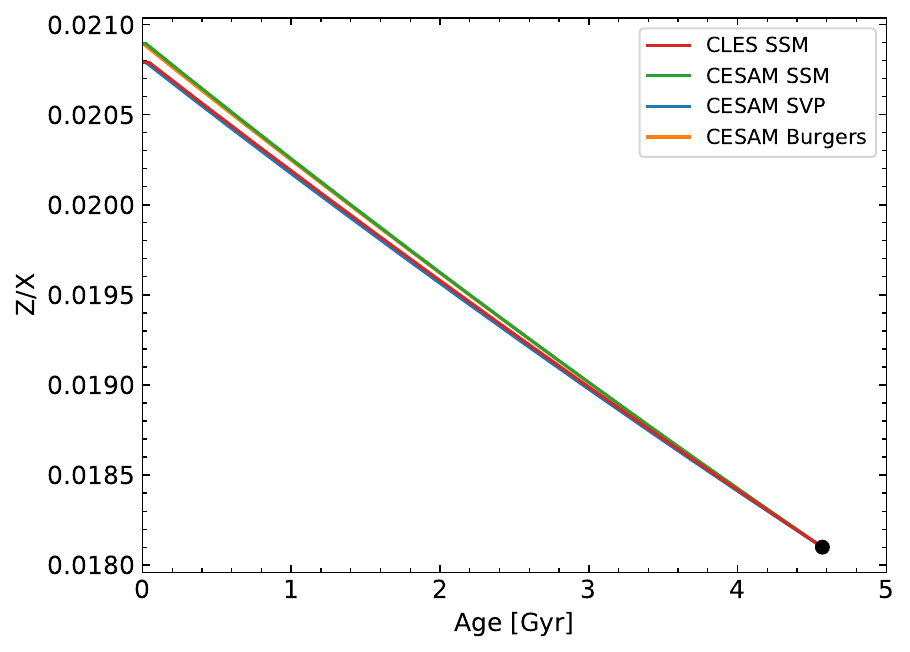}}
{\includegraphics[width=0.49\textwidth,clip=]{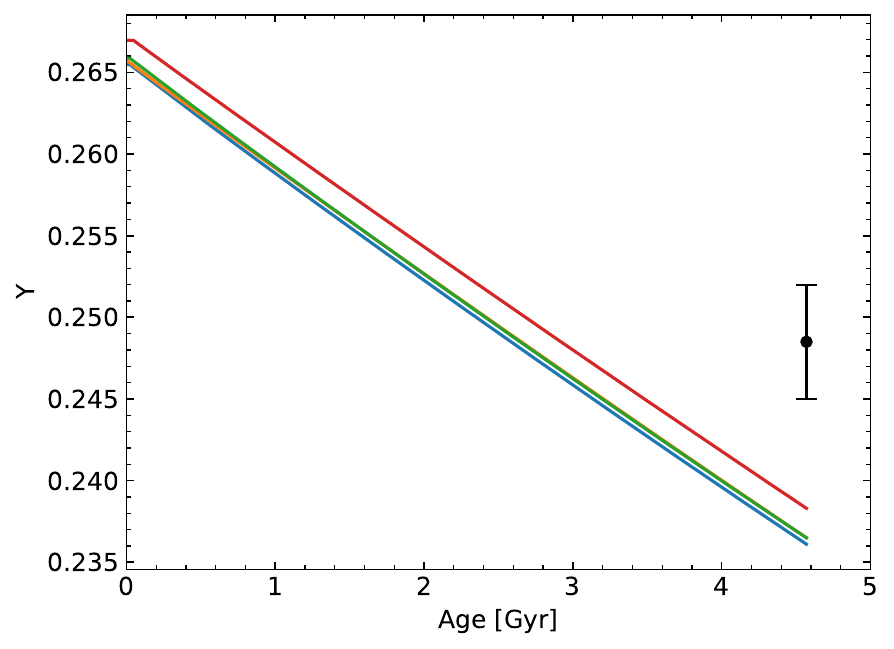}}
\centerline{\includegraphics[width=0.5\textwidth,clip=]{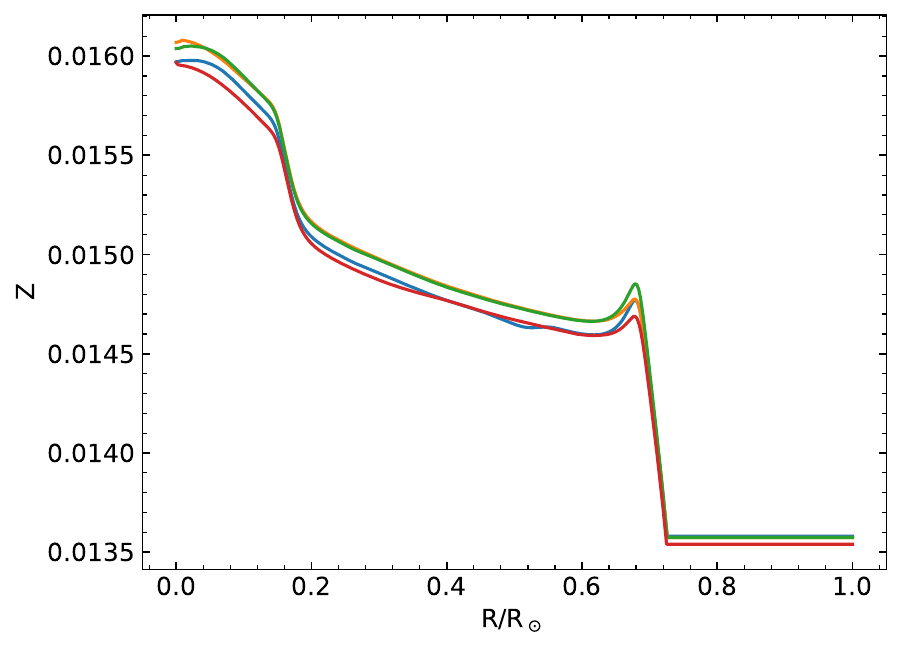}}
\small
        \caption{Evolution of the surface $Z/X$ and $Y$ with time (top panel) and $Z$ profiles according to the radius (bottom panel) for models CLES SSM and CESAM SSM, Burgers, SVP. The black symbols and error bar represent the solar values.}
\label{fig:diff}
\end{figure}

Figure~\ref{fig:diff} shows the variations of the surface $Z/X$ and $Y$, and the metallicity profiles for CLES and \cesamxx standard solar models (SSM). The variations of the surface abundances are very similar for both SSMs, with a larger initial helium abundance for the CLES model (similarly to the model without diffusion, see Sec.~\ref{sec:nodiff}). As imposed by the calibration, the final $Z/X$ is the same for both models. As expected for SSMs with low-metallicity metal mixtures, the helium surface abundance is smaller than the solar one, as already shown in multiple studies \citep[][and references therein]{Basu2008}. Although the metallicity profiles have similar behavior, the details show differences, such as a slightly lower bump at the bottom of the convective envelope and a slightly lower metallicity in the core for the CLES model. These differences come from a different treatment of atomic diffusion (see Sec.~\ref{sec:diff}) and may also be due to the treatment of the diffusion equation (instantaneous mixing in convective zones for CLES compared to diffusive convective mixing in \cesamxx). These slight differences also show in the sound speed inversions (Fig.~\ref{fig:c2DiffMod}) both in the core and at the bottom of the convective envelope. The larger difference in the core comes from differences in the temperature gradient, which seems to come from the intrinsic differences of the models described in Sec.~\ref{sec:diff} and not from the treatment of atomic diffusion. The impact of difference of the atomic diffusion formalisms is also tested within \cesamxx. CESAM SSM and Burgers models show very similar surface abundance variations and metallicity profiles (Fig.~\ref{fig:diff}). The small differences are not significant in the sound speed profiles as shown in Fig.~\ref{fig:c2DiffMod}. 

Despite the fact that radiative accelerations induce a second-order effect compared to gravitational settling for stellar models with thick convective envelopes (including the Sun), it is important to test it against the high-precision solar data. The effect of radiative acceleration can be seen with the CESAM SVP model in Figs.~\ref{fig:diff} and \ref{fig:c2DiffMod}. The depletion of metals is slightly reduced, leading to small differences in the metallicity profile that are not seen in the inversion of the sound speed. However, since radiative accelerations affect the metal mixture and therefore the Rosseland mean opacity, any potential revision of the opacity tables could lead to a change in the efficiency of this process and a change in its contribution to the opacity at the bottom of the convective envelope \citep{turcotte98,schlattl02,gorshkov08,buldgen2025b}.

From these model comparisons, we see that while atomic diffusion is crucial for solar modelling, the way it is implemented in stellar evolution codes (within the formalism tested in this work) has a minor impact on the quality of the solar models with respect to the sound speed profile. For more physically driven solar models, following the atomic diffusion for individual elements seems preferable, especially when including radiative accelerations.

\section{Additional transport and light elements}

\begin{figure}
{\includegraphics[width=0.5\textwidth,clip=]{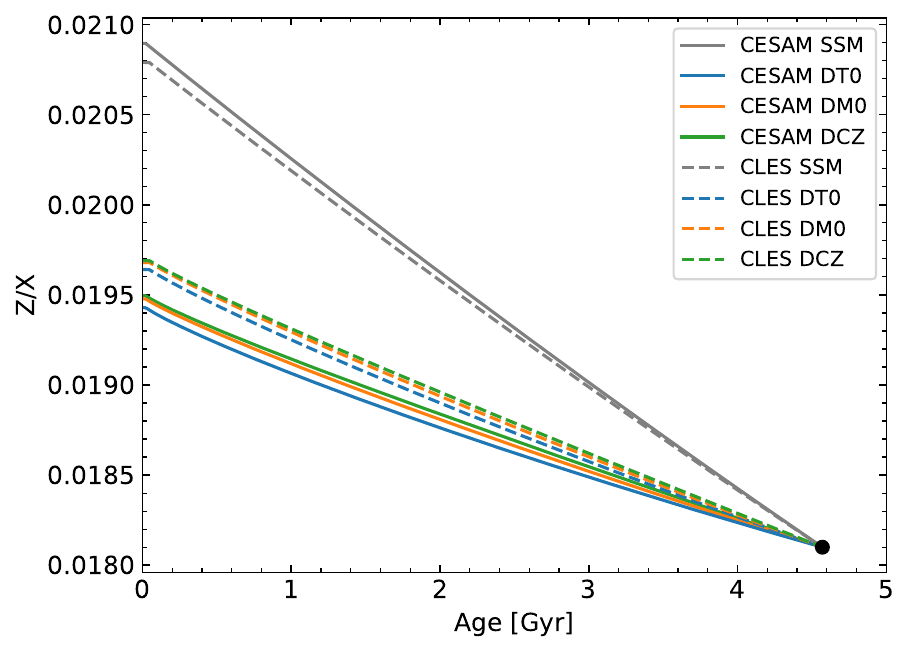}}
{\includegraphics[width=0.49\textwidth,clip=]{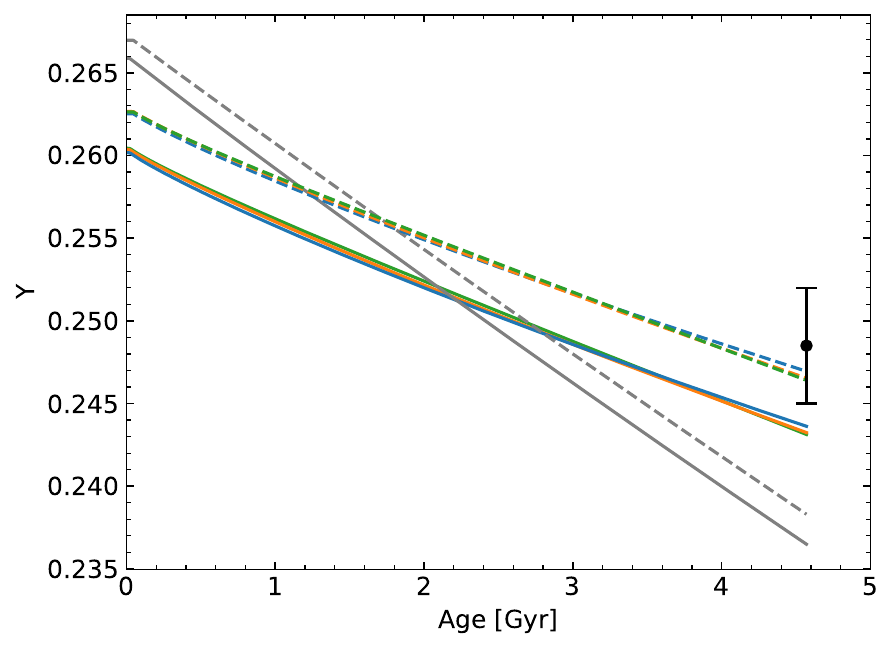}}
\centerline{\includegraphics[width=0.5\textwidth,clip=]{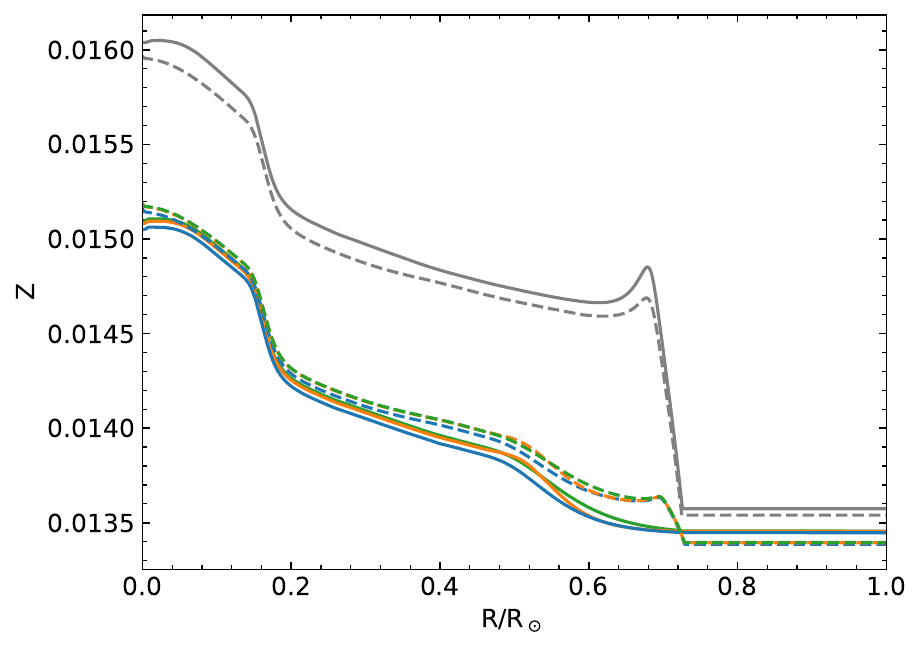}}
\small
        \caption{Same legend as Fig.~\ref{fig:diff} for models CLES (dashed lines) and CESAM (solid lines) SSM, DT0, DM0, and DCZ models. }
\label{fig:turb}
\end{figure}

Light elements such as lithium ($^7$Li) and beryllium ($^9$Be) are key ingredients in constraining the transport of chemical elements in stellar models. The Sun shows a depletion of a factor of about $100$ between its initial and current surface abundance that cannot be explained by SSMs. The only way to explain this depletion is to include additional macroscopic transport processes \citep[see e.g.][]{lebreton87}, and we know some of them are acting in the Sun. For example, hydrodynamical instabilities are driven by the rotation of the Sun (and stars in general) and transport angular momentum and chemical elements. The combination of meridional circulation and shear instability induces a transport of chemical elements that depletes lithium and beryllium. However, this depletion is orders of magnitude too large compared to observation \citep[see e.g.][]{eggenberger22}. Magneto-hydrodynamical instabilities also play a role in affecting the transport of angular momentum, hence the transport of chemicals by the effect of the shear instability \citep{maeder09}. Currently, none of the combinations of (M)HD instabilities tested in solar models was able to reproduce the surface abundances of light elements.

Accurate modelling of macroscopic transport processes responsible for the missing transport being out of the scope of this paper, we model the total macroscopic transport of chemical elements (except convection) with an ad hoc turbulent diffusion coefficient. The various expressions are described in Sec.~\ref{sec:turb}. For these models, variations of surface abundances are very different from the SSMs (see top panels of Fig.~\ref{fig:turb}). The additional mixing counteracts the effect of atomic diffusion, leading to a slower depletion of metals and helium. The calibrated solar models then start with smaller initial abundances for these elements. We note that with the additional mixing, the predicted helium abundance is closer (or within the error bar for CLES models) to the observed value.

The metallicity profiles are also affected (bottom panel of Fig.~\ref{fig:turb}), with a smaller amount of metals in the radiative zone. The difference in the treatment of the diffusion equation between both codes can be seen at the bottom of the convective envelope with a smoother variation for the \cesamxx models. The remaining metallicity bump for the CLES models is an artifact of instantaneous mixing in convective zones.

\begin{figure}
{\includegraphics[width=0.49\textwidth,clip=]{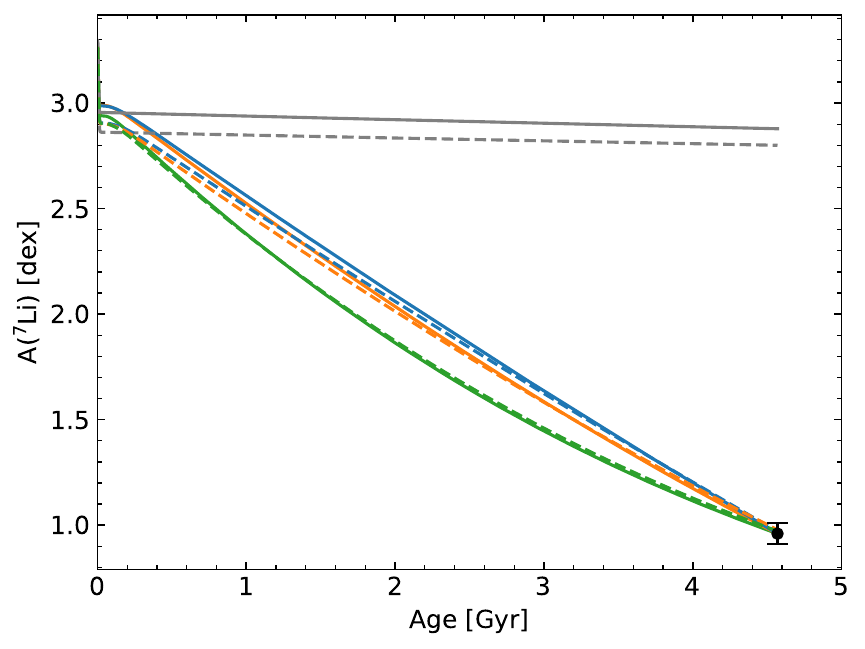}}
{\includegraphics[width=0.5\textwidth,clip=]{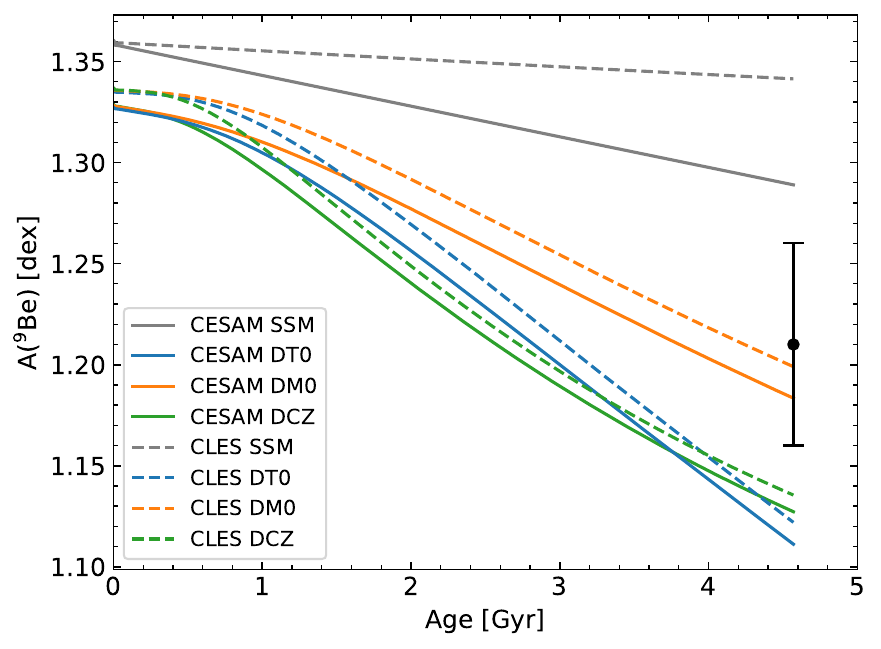}}
\small
        \caption{Evolution of $^7$Li and $^9$Be surface abundances with time for CLES (dashed lines) and CESAM (solid lines) SSMs, DT0, DM0, and DCZ models. The black symbols represent the solar values from \cite{wang21} and \cite{amarsi24} for lithium and beryllium, respectively. }
\label{fig:LiBe}
\end{figure}

The variation of the surface abundances of lithium and beryllium is presented in Fig.~\ref{fig:LiBe}. As expected from the calibration, lithium is well reproduced in all models, with slightly different behavior during the evolution. Beryllium is only well reproduced in DM0 models that are characterized by a sharper decrease in the efficiency of turbulent mixing with depth ($n=4$). This effect was already identified in previous works \citep[see e.g.][]{buldgen25}. We note that for SSMs, beryllium is depleted faster in the \cesamxx models, which seems to come from the fact that the elements are followed individually for atomic diffusion. Moreover, the differences in the initial abundances are only due to different size of convective envelope that leads to different depletion on the PMS.

\section{Electronic screening and neutrino fluxes }

As mentioned in Sect. \ref{Sec:Screening}, eletronic screening has been discussed in numerous publications and as there is no clear consensus on the question so far, it is worth investigating its impact on the measurement properties of the solar core. The best probes we have to measure the properties of the core are neutrino fluxes that have now been measured for multiple nuclear reactions \citep[see e.g.][and references therein]{Villante2021}. A detailed study using CLES models was also performed by \citet{Salmon2021}.

\begin{figure}
\centerline{\includegraphics[width=0.65\textwidth,clip=]{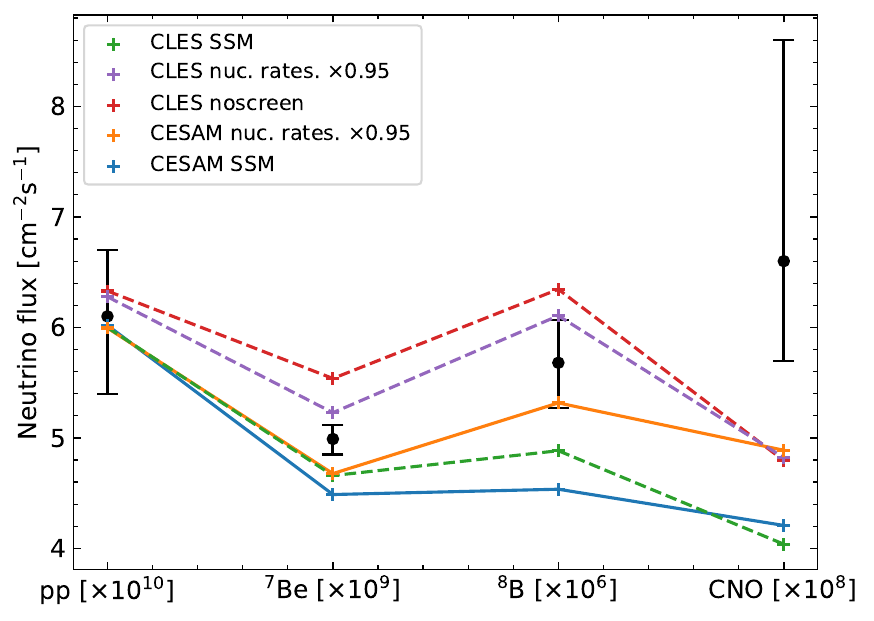}}
\small
        \caption{Fluxes for pp chain, $^7$Be, $^9$B, and CNO neutrinos for SSM, \textit{noscreen} and \textit{nuc} models. Black symbols represents the observed values from \cite{borexino22}.}
\label{fig:neut}
\end{figure}

In Fig. \ref{fig:neut}, we illustrate the neutrino fluxes for \cesamxx and CLES models including variations in the modelling of their core properties and compare them to the observed values \citep{borexino22}. These results do not serve the purpose of quantifiying the agreement or the disagreement of solar models with experimental measurements but rather to show the impact of potential revisions of physical processes may have on such comparisons. We use our SSMs as references and display the impact of a 5$\%$ modification of the pp chain reaction on the neutrino fluxes. This reduction was described by \cite{mussack11} as one way to simulate the effects of dynamical screening by the electrons for the pp chain. We show in Fig. \ref{fig:neut} that this effect is significant for all fluxes.

First, we may note that the slight differences in the core properties of the CLES and \cesamxx models (see Table~\ref{tab:core}) lead to slightly different neutrino fluxes. The differences for $^{7}$Be and $^{8}$B are the most notable one and show the sensitivity of these fluxes to the temperature gradients in the core layers. The slightly steeper gradients in the CLES models leads to higher values for these fluxes, while the pp flux is identical, as a result of the luminosity constraint in the solar calibration. Slight differences in the CNO flux are also observed and likely originate from differences in the metallicity profile in the core, with \cesamxx models having a slightly higher $Z_\mathrm{c}$ value. 

The trends observed when altering the nuclear reaction rates are similar for both codes. At the current solar age, $T_\mathrm{c}$, $\rho_\mathrm{c}$ and $Z_\mathrm{c}$ are increased as a natural effect of the lower efficiency of the pp chain while the luminosity to be reproduced has been kept the same. This leads to naturally higher  model predicted neutrino fluxes for all neutrino sources and shows the significant impact dynamical screening may have on the predictions of solar models. It also seems that these deviations seem to be amplified by the differences in temperature gradient in the core between the two codes, as pp and CNO neutrinos show the same trend, while the $^{7}$Be and $^{8}$B fluxes are quite different. As an additional test, we ran a CLES model where nuclear screening was fully deactivated for all nuclear reactions, the effect is similar to the one observed in the CLES model for which the efficiency of the pp chain has been reduced. The effect appears to be magnified even further in this case, as more reactions are modified. Both tests with reduced efficiency and deactivated electronic screening are however highly prospective and just show the potential impact it may have on our conclusions. They 
remain an important source of uncertainty in solar modelling and the interplay between the existing differences in the code and the modifications of the efficiency of nuclear reactions also calls for further analyses that are beyond the goal of this paper. Overall, the agreement between the two codes remains good and would lead to similar conclusions regarding the disagreement of solar models using the AGSS09 abundances and neutrino fluxes that can be found in the literature \citep{Serenelli2009}.

\begin{table}
\caption{Core properties properties of the models
}
\label{tab:core}
\begin{tabular}{lcccc}     
\hline                     
Model & $X_c$ & $Z_c$ & $T_c$ [$10^7$ K] & $\rho_c$ \\
\hline
CESAM nodiff & $0.384$ & $0.01365$ & $1.531$ & $146.06$   \\
CLES nodiff & $0.382$ & $0.01362$ & $1.529$ & $145.52$  \\
CESAM SSM & $0.358$ & $0.01604$ & $1.556$ & $150.81$  \\
CLES SSM & $0.359$ & $0.01597$ & $1.552$ & $149.20$   \\
CESAM Burgers & $0.358$ & $0.01607$  & $1.556$ & $151.10$  \\
CESAM SVP & $0.359$ & $0.01597$  & $1.555$ & $150.75$  \\
CESAM DT0 & $0.366$ & $0.01505$ & $1.548$ & $150.07$  \\
CLES DT0 & $0.365$ & $0.01516$ & $1.545$ & $148.52$  \\
CESAM DM0 & $0.365$ & $0.01508$ & $1.548$ & $150.10$  \\
CLES DM0 & $0.365$ & $0.01519$ & $1.545$ & $148.56$ \\
CESAM DCZ & $0.365$ & $0.01509$ & $1.548$ & $150.11$ \\
CLES DCZ & $0.364$ & $0.01518$ & $1.545$ & $148.65$  \\
CESAM nuc & $0.357$ & $0.01631$ & $1.565$ & $153.40$ \\
CLES nuc & $0.348$ & $0.01606$ & $1.566$ & $155.60$  \\
CLES noscreen & $0.361$ & $0.01607$ & $1.561$ & $152.31$  \\
CESAM CGM & $0.358$ & $0.01604$ & $1.556$ & $150.79$ \\
CESAM SSM OPAL & $0.359$ & $0.01621$ & $1.555$ & $150.71$ \\
CESAM ECM OPAL & $0.360$ & $0.01603$ & $1.554$ & $150.51$ \\
\hline
\end{tabular}
\end{table}

\section{Inversion results}\label{Sec:InvRes}

The inversions of solar structure using each of our models as reference have been performed using the SOLA inversion technique \citep{Pijpers1994} and the trade-off parameter calibration procedure of \cite{Rabello1999}. Additional details on the procedure and results on previous sets of models have been presented in \citet{Buldgen2017A, Buldgen2019,buldgen23}.

\begin{figure}
\centering{\includegraphics[width=0.8\textwidth,clip=]{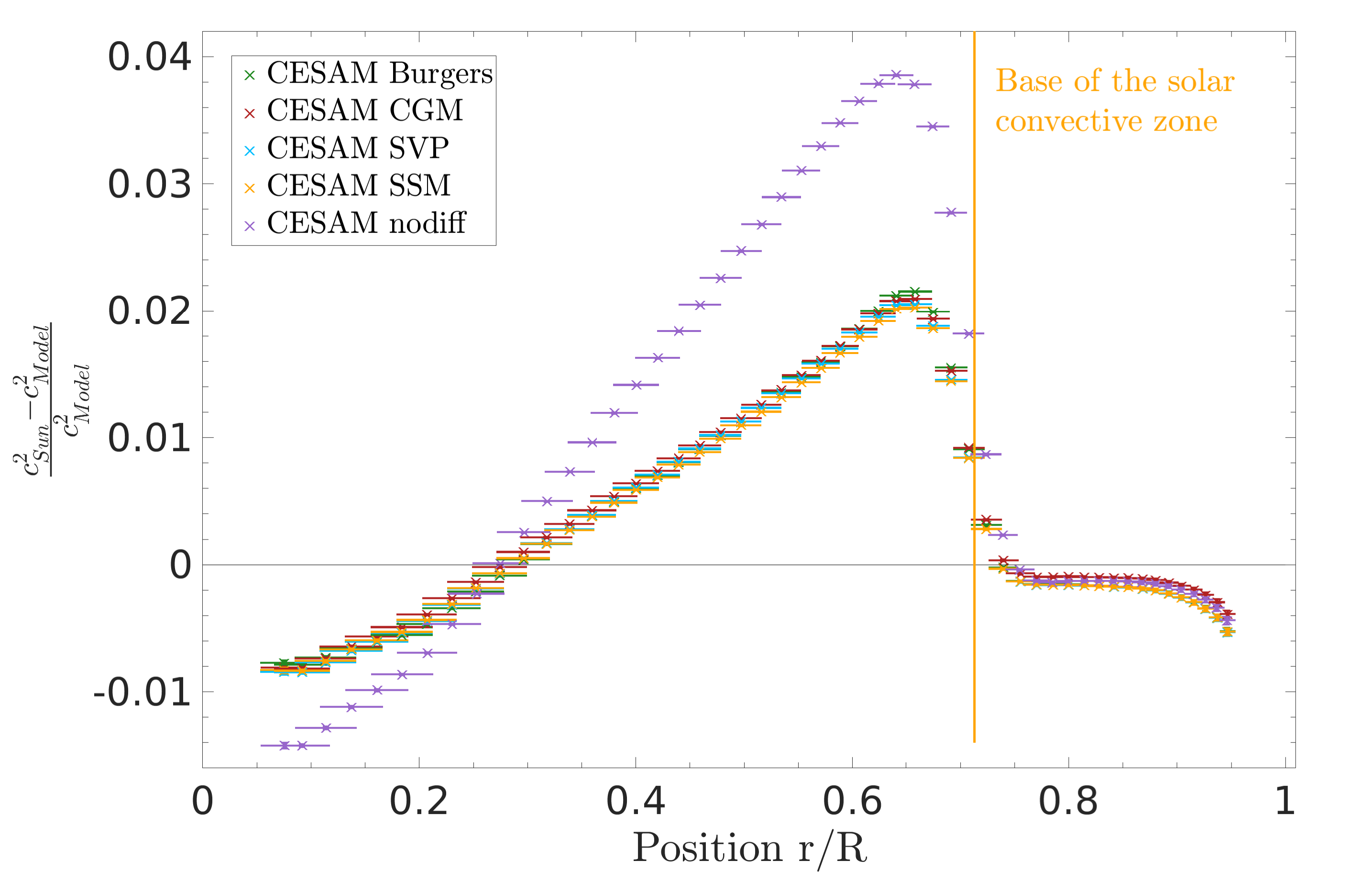}}
\small
        \caption{Relative differences between the Sun and the model squared adiabatic sound speed profile $c^{2}$ as a function of normalized radius $(r/R)$ for \cesamxx models with various implementations of microscopic diffusion.}
\label{fig:c2DiffMod}
\end{figure}

\begin{figure}
\centering{\includegraphics[width=0.8\textwidth,clip=]{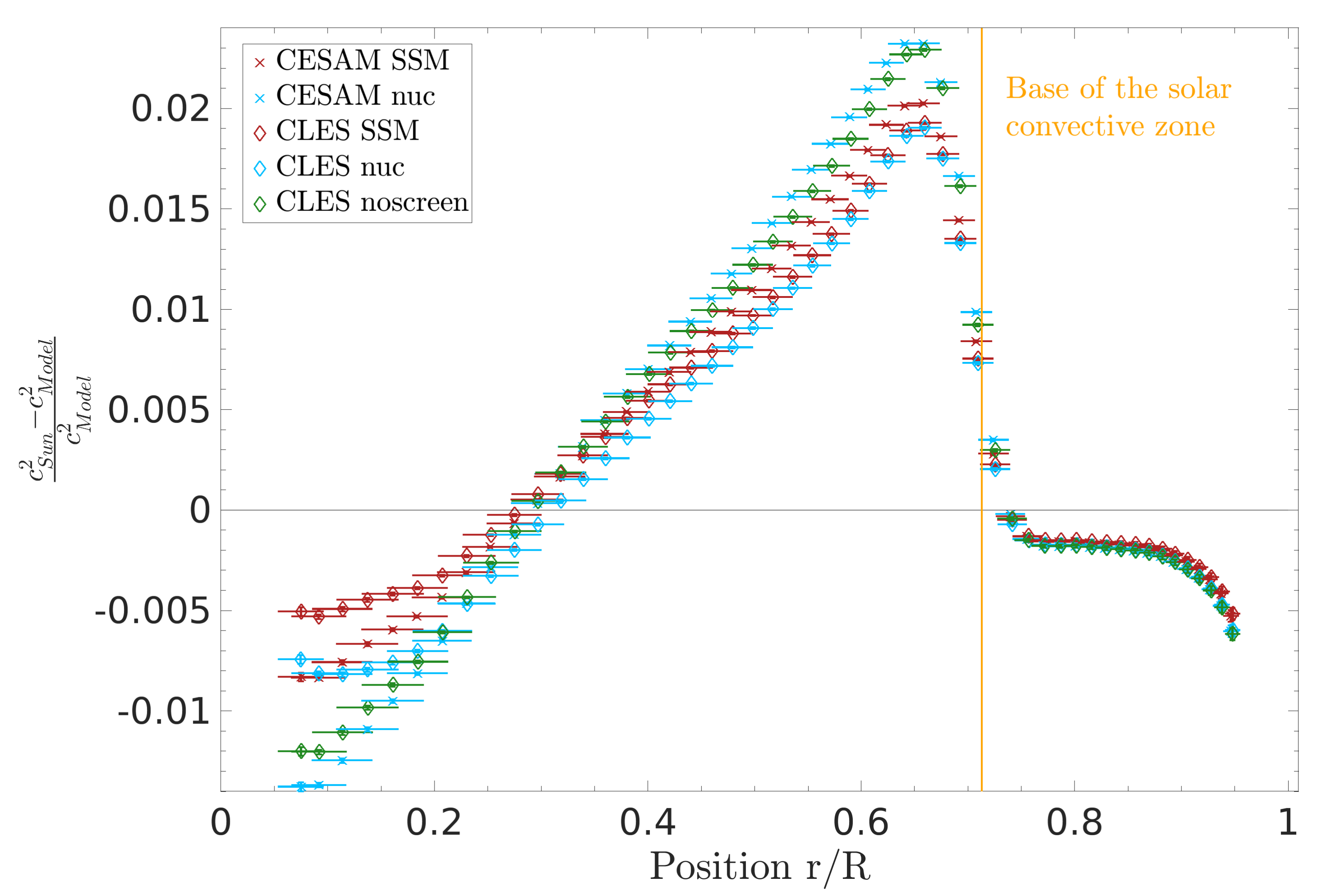}}
\small
        \caption{Relative differences between the Sun and the model squared adiabatic sound speed profile $c^{2}$ as a function of normalized radius $(r/R)$ for \cesamxx and CLES models including modifications of nuclear reactions (turning off electronic screening and pp chain efficiency to mimic the effects of dynamic screening).}
\label{fig:c2Nuc}
\end{figure}

\begin{figure}
\centering{\includegraphics[width=0.8\textwidth,clip=]{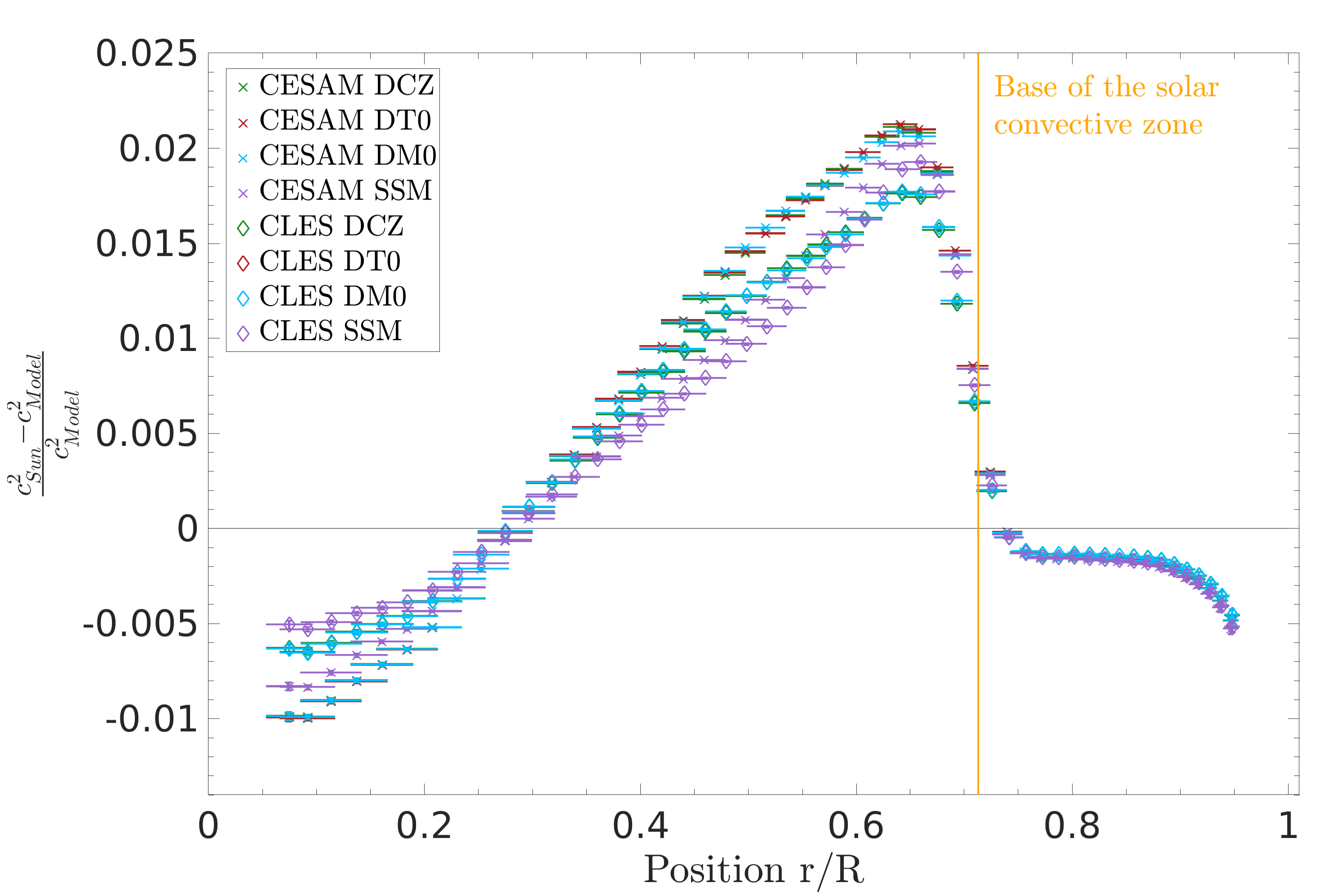}}
\small
        \caption{Differences of the squared adiabatic sound speed profile $c^{2}$ as a function of normalized radius $(r/R)$ for \cesamxx and CLES models including empirical turbulent transport.}
\label{fig:c2DT}
\end{figure}

We start by presenting the sound speed inversions for the various models, in Figs. \ref{fig:c2DiffMod}, \ref{fig:c2Nuc} and \ref{fig:c2DT}.
From Fig.~\ref{fig:c2DiffMod}, we confirm that including microscopic diffusion provides a significant improvement to the agreement of solar models with helioseismology, as already shown by \citet{JCD1993}. Furthermore, it appears that there are no large differences between the various formalisms used to describe microscopic diffusion in \cesamxx regarding the internal sound speed profile of solar-calibrated models. This seems to indicate that none of these formalisms is substantially better than the others, while physically the most complete description of microscopic diffusion is clearly the one of the CESAM SVP model.

In Figure \ref{fig:c2Nuc}, we illustrate the effect of turning off nuclear screening in the CLES models, denoted ``noscreen'' and reducing the efficiency of the pp chain reaction by 5$\%$ for both CLES and \cesamxx models, denoted ``nuc''. Compared to the standard solar models, reducing the pp chain efficiency as suggested by \citet{mussack11} leads to a worse agreement with helioseismic data, while they actually saw the opposite effect in their studies. Comparing the \cesamxx (red crosses) and CLES models (red lozenges), we can see intrinsic difference between the two SSMs, of the order of $0.1\%$ at the BCZ and $0.3\%$ in the core. These differences were already noted in the models without diffusion and are likely due to difference in temperature gradients that slightly alter the results of calibrations. They are preserved when changing the efficiency of the pp chain, but the change in nuclear reaction efficiency leads to a different response in CLES and \cesamxx. Comparing the CESAM SSM (red crosses) and CESAM nuc (blue crosses) models, we see that the sound speed profile is also impacted in the BCZ layers (of about $0.4\%$), while for the CLES models (red and blue lozenges) the changes are only localized in the core. In the core regions, both ``nuc'' models see a relative shift in sound speed of similar amplitude, by about $0.3\%$, while the models without any nuclear screening from CLES sees changes throughout the sound speed profile of about $0.6\%$. This implies that improving the neutrino fluxes using nuclear screening comes at the expense of a worsening of the agreement in sound speed.

We first note intrinsic differences between the \cesamxx and CLES models, at the level of $0.1\%$ at the BCZ and $0.3\%$ in the core. These differences are likely due to the differences in temperature gradient observed already in the models without diffusion that lead to differences in the solar calibrations. Regarding the models with modified nuclear reactions, we note that the sound speed profile at the BCZ is unchanged for the CLES model while it clearly worsens for \cesamxx models. 

The impact of including macroscopic mixing at the BCZ is illustrated in Fig. \ref{fig:c2DT}. We can see that compared to a Standard Solar Model, the improvement is not significant, as the main deviation in the sound speed profile is due to temperature gradients which are only slightly affected by the inclusion of additional mixing at the BCZ. Again, both CLES and \cesamxx models show small differences of about $0.2\%$, this time throughout the structure. The overall shape of the sound speed profile inversion is similar to what has been seen in previous studies \citep{gough96, basu97,Richard1996,Richard1997,Gabriel1997,Brun2002,JCD2018,Buldgen2019}. It also appears that all three mixing formalisms lead to very similar profiles, implying that, as long as it is included and reproduces the observed depletion, the sound speed profile is not sensitive to the exact shape of the transport coefficient.

\begin{figure}
\centering{\includegraphics[width=0.8\textwidth,clip=]{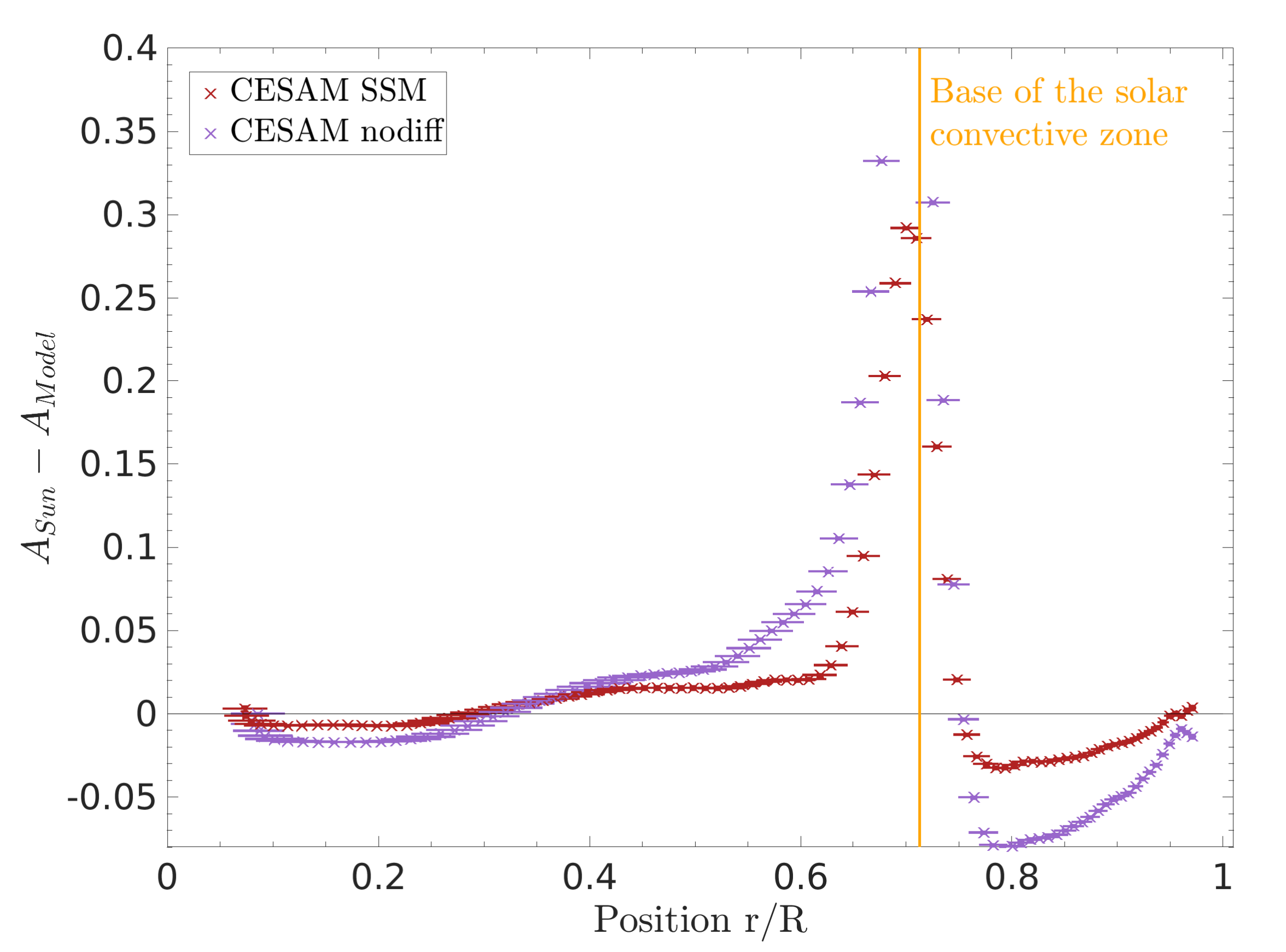}}
\small
        \caption{Differences of the Ledoux discriminant as a function of normalized radius $(r/R)$ for \cesamxx and CLES SSMs and \textit{nodiff} models.}
\label{fig:ADiff}
\end{figure}

\begin{figure}
\centering{\includegraphics[width=0.8\textwidth,clip=]{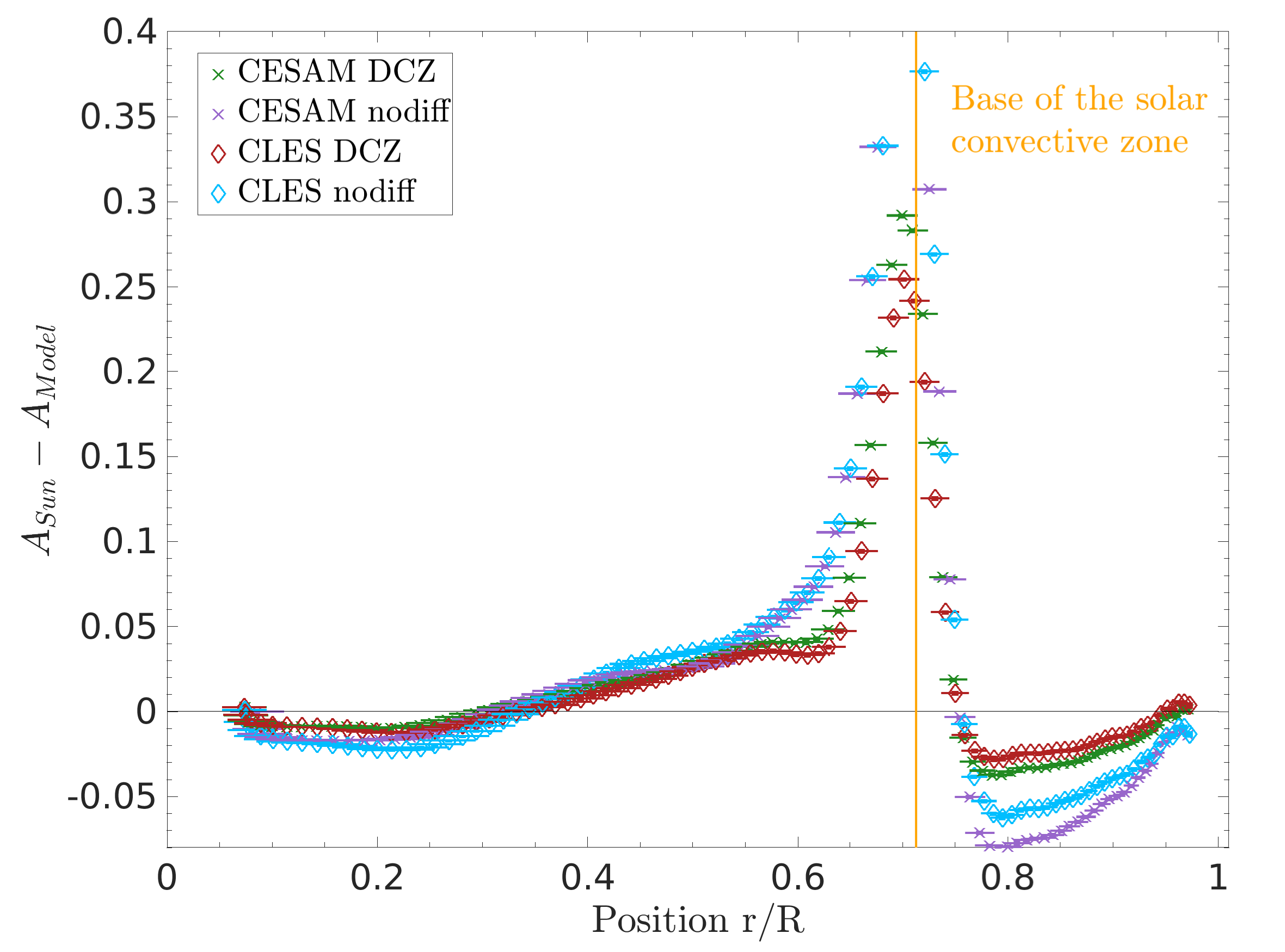}}
\small
        \caption{Differences of the Ledoux discriminant as a function of normalized radius $(r/R)$ for \cesamxx and CLES models including empirical turbulent transport.}
\label{fig:ADT}
\end{figure}

\begin{figure}
\centering{\includegraphics[width=0.8\textwidth,clip=]{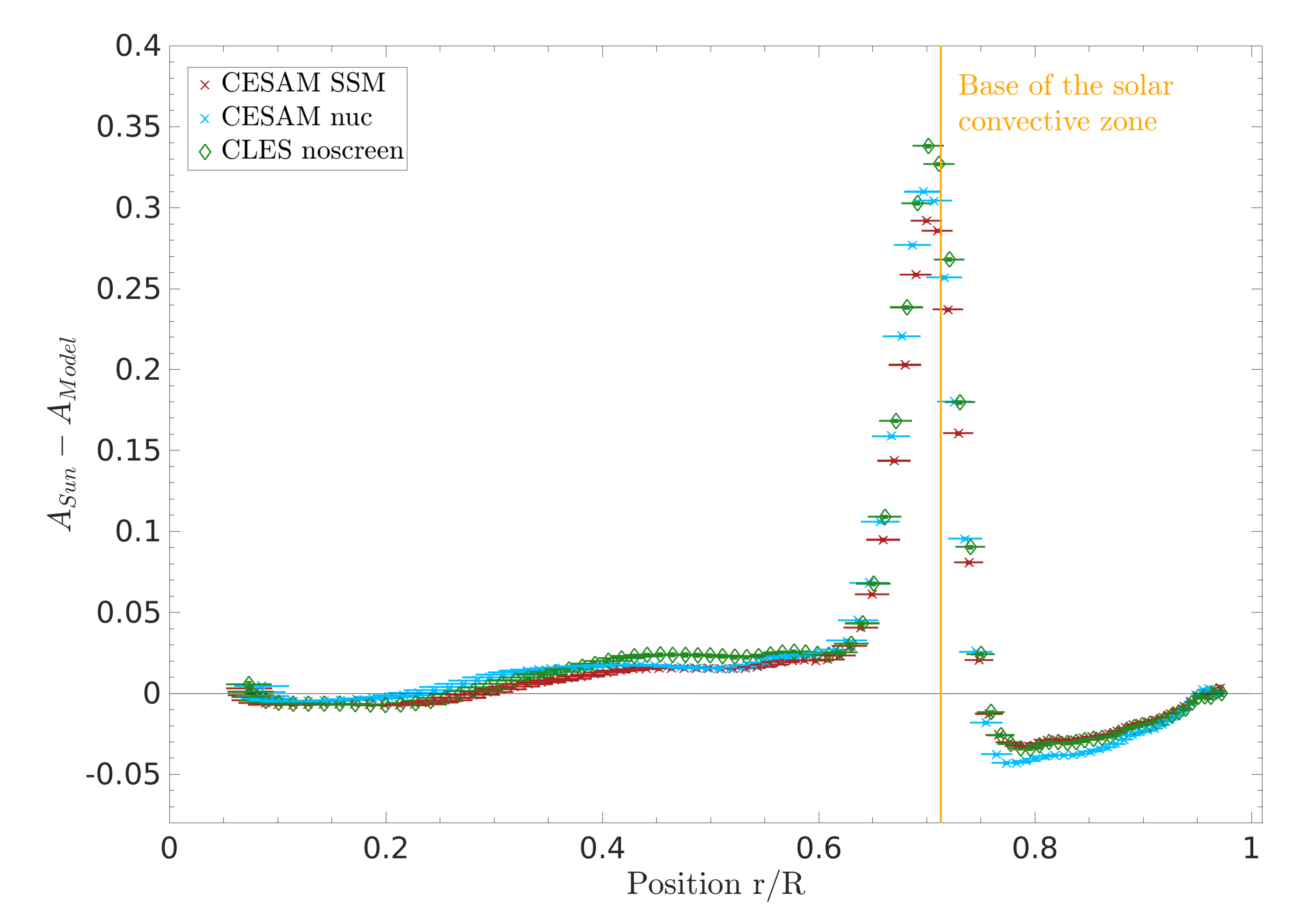}}
\small
        \caption{Differences of the Ledoux discriminant as a function of normalized radius $(r/R)$ for \cesamxx and CLES models reducing the nuclear rates or ignoring the nuclear electronic screening.}
\label{fig:ANuc}
\end{figure}

These findings are confirmed from the Ledoux discriminant inversions, that we present in Figs. \ref{fig:ADiff}, \ref{fig:ADT} and \ref{fig:ANuc}. Again it appears that microscopic diffusion significantly improves the agreement with helioseismic data, both at the BCZ and in the core regions. From Fig. \ref{fig:ADT}, we also see that a small improvement of the agreement is made when including turbulence at the BCZ, as noted by \citet{Buldgen2017A}, implying that the remaining differences are due to differences in thermal gradients at the BCZ. We also see that the agreement between the \cesamxx and the CLES models is quite good, with a slightly better agreement for the CLES models that is likely due to the slightly deeper convective envelope leading to steeper temperature gradients favoured by helioseismic data. When looking at the models with modified nuclear reactions, we see that the modifications remain limited throughout the profile. This seems to imply that the gradients are not too much affected by the change in nuclear reaction efficiency. The variations are mostly localised between $0.2$ and $0.3\rm{R_{\odot}}$, which makes sense considering that this is the region at the border of the core where the modifications would be most impactful. Close to the BCZ, the modifications are minimal, and the small differences occur from the slight shift in the position of the BCZ. Similarly to the sound speed inversions, the model with the deactivation of the electronic screening has larger discrepancies, confirming that including a detailed treatment of dynamical screening may have a significant impact on helioseismic results.

\begin{figure}
\centering{\includegraphics[width=0.8\textwidth,clip=]{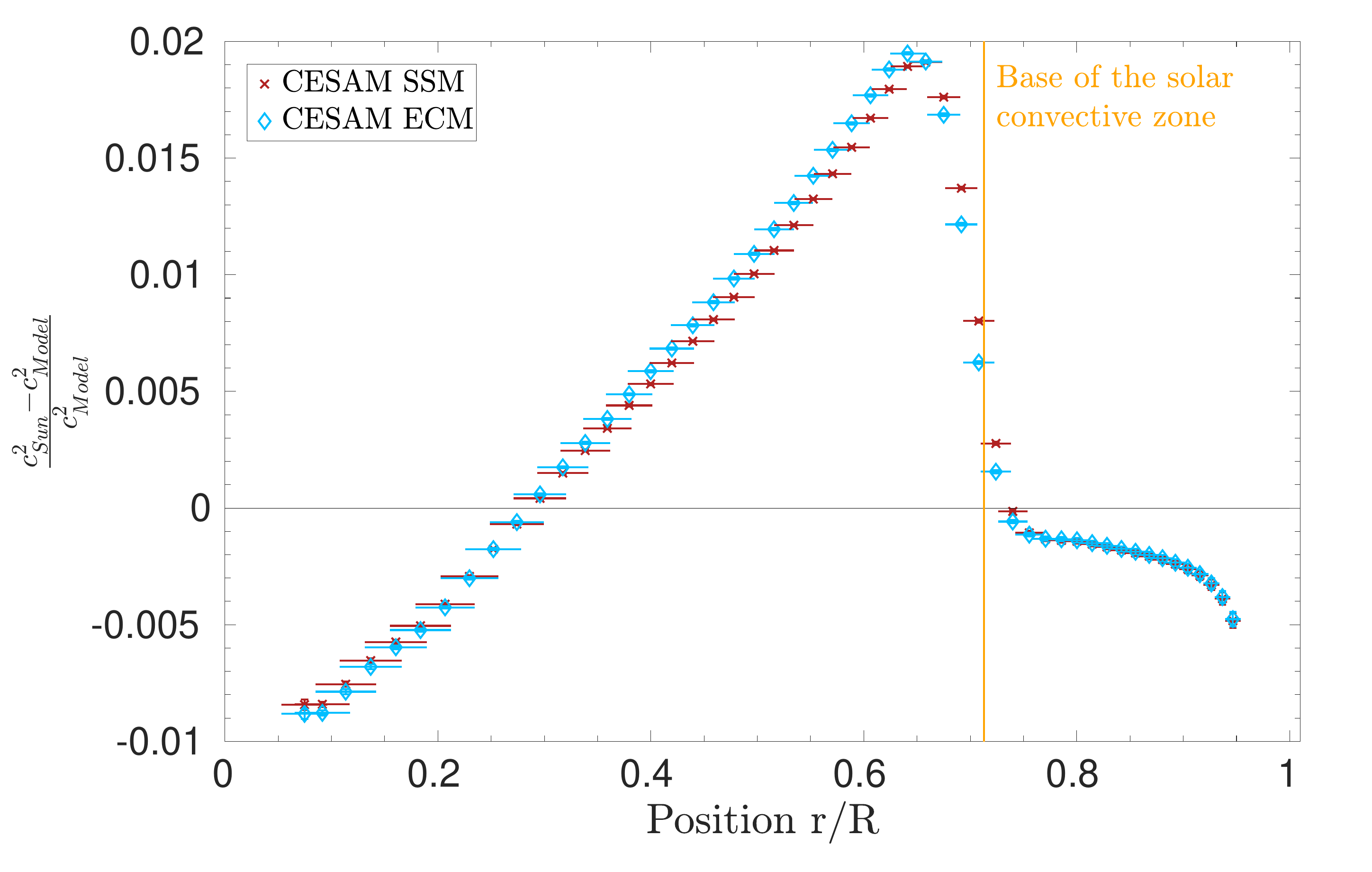}}
\small
        \caption{Relative differences of the squared adiabatic sound speed profile $c^{2}$ as a function of normalized radius $(r/R)$ for \cesamxx models using either the MLT or the ECM treatment of convection.}
\label{fig:c2ECM}
\end{figure}

The last test we computed was to compare the calibration results using the classical Mixing Length Theory to the ones using the Entropy Calibrated Modelling of \citet{manchon24}. As expected, the results presented in Fig.~\ref{fig:c2ECM} are quite similar to those obtained when comparing the CESAM SSM Model to the CESAM CGM model in Fig. \ref{fig:c2DiffMod}, as the overall impact of the change of convection formalism is negligible compared to other processes. This does not mean that the modelling of convection is negligible for solar models, as it is likely that an inversion of $\Gamma_{1}$ would exhibit large differences in the outer layers that would be significant for other helioseismic analyses, such as the determination of the chemical composition of the solar envelope or the equation of state of the solar plasma \citep{Vorontsov2013, Vorontsov2014, Buldgen2024a, Baturin2024,Baturin2025}.

\section{Conclusion}

We compared solar models computed with CLES and \cesamxx using different input physics in order to assess the impact of different ingredients on solar modelling. Both codes are in good agreement for all the configurations tested in this work. In addition to the physical effects of transport processes and nuclear screening, we identified that the implementation of these processes can have a direct effect on the modelling. It is not visible when modelling distant stars, but reveals itself when looking at the exquisite quality solar data. While both codes compare very well in many respects, the slight differences in the preparation and interpolation of opacity tables lead to slight differences in temperature gradients and properties which follow from them (e.g. the position of the base of convective envelopes and the neutrino fluxes). In addition, the implementation of atomic diffusion and the different treatments of the diffusion equation are key as they have an impact on the abundance profiles at the age of the Sun. Despite these slight differences, CLES and \cesamxx provide very similar solar models that are a good basis for assessing the impact of physical processes.

We showed that different formalisms to evaluate the microscopic diffusion velocities of chemical elements (namely T94, MP93 and B69) give very similar results and inversions of sound speed or Ledoux discriminant do not favor one over the other. The fact that atomic diffusion (and the transport of chemical elements in general) is solved for individual elements in \cesamxx and for a mean metal in CLES has no significant impact on solar models. The same applies to instantaneous mixing in convective zones in CLES and diffusive mixing in \cesamxx. However, we believe that the approach of \cesamxx for both aspects should be favored as it may treat more consistently situations where instantaneous mixing may not be physical (e.g. when the timescale of nuclear reactions is short enough to be comparable to the mixing timescale). 

SSMs are proved to fail to reproduce the abundances of lithium and beryllium, which is why additional transport processes are needed. We confirmed previous results showing that the addition of turbulent mixing allows one to reproduce the surface abundances of both elements but degrades the agreement for inversions. A sharp decrease in mixing efficiency with density ($n=4$ in Eq.~\ref{eq:dturb1} and \ref{eq:dturb2})is also favored to specifically reproduce the abundance of beryllium, as pointed out in previous studies \citep[e.g.][]{buldgen2025b}.

We showed that treatment of convection has a very small impact on the sound speed inversion, as it leads to minor shifts in the radiative zone. This does not mean that the modelling of convection will not have an impact on helioseismology of the solar envelope, but it is not surprising that its impact on the sound speed profile in the radiative zone remains limited compared to other physical processes. 

Reducing the efficiency of the nuclear reaction rates (weak screening included) of $5$\% as suggested by \cite{mussack11} and \cite{Dappen2024} has the same effect as neglecting the electronic screening of nuclear reactions on the core properties. It greatly affects the predictions of neutrino fluxes, but is still far from the observations of the CNO neutrinos. This confirms that electronic screening of nuclear reactions is an important ingredient for accurate solar models.

The thorough comparisons between solar calibrated models presented here have shown how small differences between stellar evolution codes may impact some conclusions drawn from detailed analyses of helioseismic constraints and neutrino fluxes. While these discrepancies remain small, especially compared to variations of physical ingredients, they are not entirely negligible at the current level of precision of observational constraints. In this respect, further comparisons may lead to mutual progress of both \cesamxx and CLES, but also a better understanding of the impact of various numerical recipes or simplifying assumptions made in the codes. Such comparisons are not timely solely for the analysis of helioseismic data, but also for the preparation of the upcoming PLATO mission \citep{rauer25}.

\section{Additional statements}
 
\begin{authorcontribution}
M. D. proposed the idea of testing different micro- and macroscopic transport of chemical element formalisms on solar models, and G. B. proposed the idea of testing the impact of electronic screening on solar models and neutrino fluxes. The idea of comparing CLES and \cesamxx in this context was decided by all co-authors. CLES models were computed by G. B., A. N., and R. S., and \cesamxx models were computed by M. D., L. M., and Y. L. The comparison of internal structures and evolution of surface abundances was performed by M. D. and inversions were performed by G. B. All co-authors contributed to the interpretation of the results and were involved in the discussions.
\end{authorcontribution}

\begin{fundinginformation}
M. D. acknowledges support from CNES, focused on PLATO. G. B. acknowledges fundings from the Fonds National de la Recherche Scientifique (FNRS) as a postdoctoral researcher. L. M. acknowledges support from the Agence Nationale de la Recherche (ANR) grant ANR-21-CE31-0018.
\end{fundinginformation}

\begin{dataavailability}
All observational data are public and accessible from the references mentioned in the text.
\end{dataavailability}


\begin{codeavailability}
The \cesamxx stellar evolution code and OSM code are publicly available and can be accessed from this website: \url{https://www.ias.u-psud.fr/cesam2k20/home.html}.
\end{codeavailability}

\begin{ethics}
\begin{conflict}
The authors declare that they have no conflicts of interest.
\end{conflict}
\end{ethics}

\bibliographystyle{spr-mp-sola}
\bibliography{transport}  

\IfFileExists{\jobname.bbl}{} {\typeout{}
\typeout{****************************************************}
\typeout{****************************************************}
\typeout{** Please run "bibtex \jobname" to obtain} \typeout{**
the bibliography and then re-run LaTeX} \typeout{** twice to fix
the references !}
\typeout{****************************************************}
\typeout{****************************************************}
\typeout{}}

\end{document}